# An investigation of avalanche noise in the non local impact ionization regime.

J. S. Marsland

Department of Electrical Engineering and Electronics

University of Liverpool

Brownlow Hill

Liverpool      L69 3GJ

United Kingdom

Telephone: (44) 151 794 4536

Fax: (44) 151 794 4540

Email: marsland@liv.ac.uk

Abstract

The probability distribution function (PDF) for avalanche multiplication and the excess noise factor are calculated using non local ionization coefficients. The dead space effect is shown to reduce the excess noise factor for a given multiplication. At higher electric field values a resonance effect is observed in addition to the dead space effect. The resonance effect leads to a more deterministic behaviour in the PDF for avalanche multiplication and this gives rise to very low excess noise factors for multiplication values of 2, 4, 8 and 16 if the ionization coefficient ratio is small.

1. Introduction

Avalanche photodiodes can amplify a signal current by the process of impact ionization where a carrier relaxes energy by creating an electron hole pair. However impact ionization is a stochastic process so that current is not multiplied uniformly - each carrier may ionize a different number of times and this also applies to the secondary carriers produced by impact ionization. The result is that additional noise, avalanche noise, is produced during avalanche multiplication. The excess noise factor, $F(M)$, is a measure of this additional noise and McIntyre [1] derived the following equation for the excess noise factor for a mean multiplication of $M$ initiated by electrons injected into the high field region of an avalanche photodiode.

$$F(M) = kM + \left\{2 - \frac{1}{M}\right\}(1-k) \qquad (1)$$

where $k$ is the ratio of the hole ionization coefficient, $\beta$, divided by the electron ionization coefficient, $\alpha$. The ionization coefficient is the average number of impact ionizations initiated by a carrier per unit distance and equation (1) assumes that $\alpha$ and $\beta$ are only functions of the electric field local to the carrier. However a carrier can only initiate impact ionization if it has sufficient kinetic energy above a certain threshold. This threshold energy is equal to or greater than the band gap energy of the semiconductor and carriers that start from zero kinetic energy must traverse a 'dead space' before they can gained the threshold energy from the electric field. The ionization coefficient must be zero in the dead space for any value of local electric field and this is a non local effect not included in the derivation of equation (1) above.

Okuto and Crowell [2] developed expressions for avalanche multiplication in the non local regime but did not consider noise. Avalanche noise including non local effects was considered by vanVliet [3], Marsland [4], Saleh et al [5], Hayat et al [6], Marsland et al [7] and Hayat et al [8]. All of these papers showed that equation (1) overestimates the excess noise factor by neglecting the dead space. The effect of the dead space is to reduce both the multiplication and the avalanche noise so that for a given multiplication the excess noise factor is less than that predicted by McIntyre [1]. This was subsequently confirmed experimentally by Hu et al [9] and numerous others e.g. Li et al [10].

Renewed interest in non local impact ionization has produced new theoretical studies by numerous workers including McIntyre [11] and Marsland [12]. In [12] a new definition is proposed for the non local ionization coefficient different from the definition used by all other workers from Okuto and Crowell [2] onwards. This leads to the identification of a new non local effect in addition to the well understood dead space effect. This 'resonance' effect occurs alongside the dead space effect at very high electric fields. The previous paper [12] was restricted to non local impact ionization and avalanche multiplication only. In this paper the probability distribution function (PDF) for avalanche multiplication and the excess noise factor are calculated at fields where both the dead space and resonance effects can be observed.

A resume of the previous paper [12] is given in Section 2. Expressions for the PDF for avalanche multiplication are derived in Section 3 and these are evaluated for the non local case. In Section 4 the excess noise factor, $F(M)$, is calculated for the non local case and very low values of $F(M)$ are predicted at very high fields. Section 5 present conclusions.

## 2. Resume of previous work.

In the previous paper [12] the non local ionization coefficient, $\alpha(z)$, is defined such that $\alpha(z)dz$ is the probability that a carrier will impact ionize in the interval $(z, z + dz)$ starting with no kinetic energy at $z = 0$. Using this definition a relationship between $\alpha(z)$ and the probability distribution function (PDF) for the pathlength to ionization, $h(z)$, can be found:

$$\alpha(z) = h(z) + \int_0^z \alpha(x) h(z-x) dx \qquad (2)$$

The ionization pathlength PDF, $h(z)$, is defined such that $h(z)dz$ is the probability that a carrier undergoes impact ionization for the first time in the interval $(z, z + dz)$ starting with no kinetic energy at $z = 0$. The ionization pathlength PDF can be found by Monte Carlo simulation [13-15] and the characteristic shape of $h(z)$ is a fast rising exponential starting immediately after the dead space followed by a slower decaying exponential. The following equation was proposed in [12] to model this behaviour.

$$h(z) = \frac{ab}{b-a}\{\exp(-a(z-l)) - \exp(-b(z-l))\}U(z-l) \qquad (3)$$

where $l$ is the length of the dead space, $a$ and $b$ are constants determined by the rise and fall of $h(z)$ and $U(z - l)$ is a unit step function equal to zero for $z < l$ and one elsewhere. Having assumed this equation for $h(z)$ it can be shown that the non local ionization coefficient, $\alpha(z)$, as defined by equation (2) is given by the following expressions.

$$\alpha(z) = \sum_{n=1}^{\infty} h_n(z) \tag{4a}$$

where

$$h_n(z) = a^n b^n U(z-nl) \sum_{m=1}^{n} \frac{(-1)^{n-m}(2n-m-1)!(z-nl)^{m-1}}{(n-1)!(n-m)!(m-1)!} \left\{ \frac{\exp(-a(z-nl))}{(b-a)^{2n-m}} + \frac{\exp(-b(z-nl))}{(a-b)^{2n-m}} \right\}$$

(4b)

The non local ionization coefficient, $\alpha(z)$, is the sum of functions, $h_n(z)$, that represents the pathlength PDF for the n$^{th}$ impact ionization. At low fields these functions merge so that $\alpha(z)$ has a dead space followed by a fast exponential rise to a steady value given by

$$\alpha(z \to \infty) = \left( \frac{1}{a} + \frac{1}{b} + l \right)^{-1} \tag{5}$$

This behaviour is demonstrated in Figure 2 of [12] where the parameters $a$, $b$, and $l$ where found by fitting equation (3) to the ionization pathlength PDF for GaAs at a field of $3 \times 10^7$ Vm$^{-1}$ determined by Monte Carlo simulation [15]. At a very high electric field the $h_n(z)$ functions become isolated so that the non local ionization coefficient consists of a series of isolated peaks as shown in Figure 3 of [12] for GaAs at a field of $10^8$ Vm$^{-1}$.

The non local ionization coefficient can be used to find the avalanche multiplication, $M(z)$, for an electron - hole (e-h) pair created at point $z$ in a multiplication region of width, $W$, using the following equation:

$$M(z) = 1 + \int_0^z \alpha(z-x)M(x)dx + \int_z^W \beta(x-z)M(x)dx \qquad (6)$$

Electrons have a non local ionization coefficient given by $\alpha(z)$ and are assumed to travel in the negative $z$ direction and holes have a non local ionization coefficient given by $\beta(z)$ and are assumed to travel in the positive $z$ direction. In the previous work [12] multiplication was calculated for electrons injected into the avalanche region with a constant field of either $3 \times 10^7$ Vm$^{-1}$ or $10^8$ Vm$^{-1}$ for a range of different multiplication region widths, $W$, and ionization coefficient ratios. The ionization coefficient ratio, $k$, is defined as $\beta(z)$ divided by $\alpha(z)$ as $z$ tends to infinity.

$$k = \frac{\beta(z \to \infty)}{\alpha(z \to \infty)} \qquad (7)$$

At the lower field value the multiplication was less than that calculated using local ionization coefficients and the difference was fully accounted for by the inclusion of the dead space. At the higher field value the multiplication, $M(W)$, plotted against multiplication region width, $W$, (Figure 9 of [12]) displays level sections at $M(W) = 2$, 4, 8 and 16. These level sections cannot be explained in terms of the dead space alone and they are connected with the isolated peaks in the non local ionization coefficient at the higher field value; the so called 'resonance' effect.

## 3. Probability distribution function of avalanche multiplication

### A. Theory

In this section a method is presented for calculating the probability, $P_m(z)$, that an e-h pair created at $z$ has a multiplication of $m$. Consider an e-h pair created in a multiplication region of width $W$ occupying the interval $0 < z < W$. Electrons are assumed to travel in the negative $z$ direction and holes in the positive $z$ direction. The ionization pathlength PDF for the electron travelling from $z$ to $x$ where $x < z$ is $h_e(z - x)$. So the probability that the electron survives to $x = 0$ without suffering impact ionization is given by $P_{Se}(z)$ where:

$$P_{Se}(z) = 1 - \int_0^z h_e(z-x)dx \tag{8}$$

Similarly for the hole travelling from $z$ to $x$ where $x > z$ the ionization pathlength PDF is $h_h(x - z)$ and the probability that the hole survives to $x = W$ without suffering impact ionization is given by $P_{Sh}(z)$ where:

$$P_{Sh}(z) = 1 - \int_z^W h_h(x-z)dx \tag{9}$$

If both the electron and the hole created at $z$ leave the multiplication region without impact ionizing then a multiplication of 1 results. Therefore the probability that the e-h pair created at $z$ has a multiplication of 1 is the product of the two survival probabilities given by equations (8) and (9).

$$P_1(z) = P_{Se}(z)P_{Sh}(z) \tag{10}$$

The probability that an electron created at $z$ impact ionizes in the interval $(x, x + dx)$ and that neither the original electron nor the new e-h pair undergo further ionizations is:

$$h_e(z-x)P_{Se}(x)P_1(x)dx \tag{11}$$

This assumes that (after the ionization) the initial electron and the e-h pair have the same ionization pathlength PDF as a carrier starting with no kinetic energy: a common assumption in previous work [2, 6, 11, 12]. The probability that an electron created at $z$ results in one and only one ionization is:

$$P_{e1}(z) = \int_0^z h_e(z-x)P_{Se}(x)P_1(x)dx \tag{12}$$

Similarly the probability that a hole created at $z$ results in one and only one ionization is:

$$P_{h1}(z) = \int_z^W h_h(x-z)P_{Sh}(x)P_1(x)dx \tag{13}$$

An e-h pair created at $z$ will result in a multiplication of 2 if the electron ionizes once only and the hole does not ionize or the hole ionizes once only and the electron does not ionize. Hence the probability of a multiplication of 2 for an e-h pair created at $z$ is:

$$P_2(z) = P_{e1}(z)P_{Sh}(z) + P_{Se}(z)P_{h1}(z) \tag{14}$$

The probability that an electron created at $z$ impact ionizes in the interval $(x, x + dx)$ and that the three resulting carriers have one (and only one) more ionization is:

$$h_e(z-x)\{P_{Se}(x)P_2(x) + P_{e1}(x)P_1(x)\}dx \tag{15}$$

So the probability that an electron created at $z$ results in two and only two ionizations is:

$$P_{e2}(z) = \int_0^z h_e(z-x)\{P_{Se}(x)P_2(x) + P_{e1}(x)P_1(x)\}dx \tag{16}$$

Likewise the probability that a hole created at $z$ results in two and only two ionizations is:

$$P_{h2}(z) = \int_z^W h_h(x-z)\{P_{Sh}(x)P_2(x) + P_{h1}(x)P_1(x)\}dx \tag{17}$$

An e-h pair created at $z$ will result in a multiplication of 3 if the electron ionizes twice and the hole does not ionize or the hole ionizes twice and the electron does not ionize or both carriers ionize once only. Hence the probability of a multiplication of 3 for an e-h pair created at $z$ is:

$$P_3(z) = P_{e2}(z)P_{Sh}(z) + P_{e1}(z)P_{h1}(z) + P_{Se}(z)P_{h2}(z) \qquad (18)$$

In general the probability, $P_{ei}(z)$, that an electron created at $z$ will result in a further $i$ impact ionizations is given by the following equation where $P_{e0}(x)$ is the survival probability, $P_{Se}(x)$.

$$P_{ei}(z) = \int_0^z h_e(z-x) \sum_{n=0}^{i-1} P_{en}(x)P_{(i-n)}(x)dx \qquad i \geq 1 \qquad (19)$$

Similarly the probability, $P_{hi}(z)$, that a hole created at $z$ will result in a further $i$ impact ionizations is given by the following equation where $P_{h0}(x)$ is the survival probability, $P_{Sh}(x)$.

$$P_{hi}(z) = \int_z^W h_h(x-z) \sum_{n=0}^{i-1} P_{hn}(x)P_{(i-n)}(x)dx \qquad i \geq 1 \qquad (20)$$

Finally the probability, $P_m(z)$, that an e-h pair created at $z$ results in a multiplication of $m$ and only $m$ is given by the following general equation.

$$P_m(z) = \sum_{n=1}^m P_{e(m-n)}(z)P_{h(n-1)}(z) \qquad m \geq 1 \qquad (21)$$

These equations (19, 20, 21) are essentially an alternative form of equations (25) and (26) of Hayat et al [6].  The probability, $P_m(W)$, that an electron injected into the high field region has a multiplication of $m$ is equivalent to the gain PDF, $P_G(k)$, derived by Hayat et al [6] where $k = m$ ($k$ as defined in [6]).

B. Results

$P_m(z)$ can be calculated for any $m$ and for any $h_e(z)$ and $h_h(z)$ by first calculating $P_{e0}(x) = P_{Se}(x)$ and $P_{h0}(x) = P_{Sh}(x)$ using equations (8) and (9) and then alternately calculating $P_m(z)$ followed by $P_{ei}(x)$ and $P_{hi}(x)$ using equations (19), (20) and (21) starting with $m = i = 1$ and then with increasing values of $m$ and $i$ until the following equation is satisfied for all $z$ within some predetermined error.

$$\sum_{m=1}^{N} P_m(z) = 1 \qquad N \to \infty \qquad (22)$$

The value of $N$ required for convergence increases with multiplication and with avalanche noise and typical values are shown in Table 1.  The mean multiplication for an e-h pair created at $z$ is given by the following equation.

$$M(z) = \sum_{m=1}^{N} m.P_m(z) \qquad N \to \infty \qquad (23)$$

In the limit as $N$ tends to infinity the value of $M(z)$ derived from equation (23) is identical to that found using equation (6) and this provides another check on the convergence.  It should be noted that equations (6) and (23) only give the same result if a non local ionization coefficient defined by equation (2) is used in equation (6) and that the equivalent ionization pathlength PDF is used in the derivation of $P_m(z)$.

Figure 1 shows $P_m(z)$ plotted against $z$ for a multiplication width, $W$, of 3.227 µm with $\beta(z) = 0$ and $\alpha(z)$ given by $a = 1$ µm$^{-1}$, $b = 10$ µm$^{-1}$ and $l = 0.15$ µm. The width, $W$, is chosen so that $M(W) \approx 8$. Each line corresponds to a different value of $m$ in the range 1 to 28 with solid lines for odd values of $m$ and dashed lines for even. $m$ increases monotonically from left to right for $P_m(z) < 0.05$. In the region $0 < z < 0.15$ µm only $P_1(z)$ is non zero because the dead space of electrons created at $z$ reaches the end of the multiplication region and the multiplication is 1 with certainty.

Figure 2 shows $P_m(W)$ plotted against $m$ for $M(W) \approx 8$ and for ionization coefficient ratios $k = 0$ (+), $k = 0.1$ (×) and $k = 1$ (o) as defined by equation (7). The parameters used for $\alpha(z)$ are the same as those used for Figure 1 and $\beta(z) = \alpha(z)$ for $k = 1$, $\beta(z) = 0$ for $k = 0$ and $\beta(z)$ is given by $a = 0.0816$ µm$^{-1}$, $b = 10$ µm$^{-1}$ and $l = 0.15$ µm for $k = 0.1$. The results are normalised to a mean multiplication of 8 by selecting the multiplication width. When $k = 1$ a multiplication of one is the most probable whereas $m = 2$ and $m = 4$ are the most probable for $k = 0.1$ and $k = 0$ respectively. $P_m(W)$ tends to zero as $m$ increases and this occurs more slowly for a larger $k$. This leads to a higher excess noise factor for larger $k$ as predicted by McIntyre [1]. The results shown in Figures 1 and 2 are broadly similar to those predicted using local ionization coefficients.

Now consider $P_m(z)$ calculated using parameters found by fitting Monte Carlo simulations [15] for GaAs at the higher field value of $10^8$ Vm$^{-1}$. Figure 3 shows $P_m(z)$ plotted against $z$ for a multiplication width, $W$, of 0.173 µm with $\beta(z) = 0$ and $\alpha(z)$ given by $a = b = 300$ µm$^{-1}$ and $l = 0.0415$ µm. Each line corresponds to a different value of $m$ in the range 1 to 8 with dashed lines for even values of $m$ and solid lines for odd. No electrons created in the region $0 < z < 0.0415$ µm can ionize due to the dead space so $P_1(z) = 1$ in this region. Electrons created at $z \approx 0.08$ µm will ionize once and only once so that $P_2(z) = 1$ at this point. At $z \approx 0.1$ µm the electron can ionize twice or alternatively it will ionize once and the secondary electron will also ionize once so that a multiplication of three is possible although no more likely than $m = 2$ or $m = 4$. An electron created at $z \approx 0.125$ µm will ionize twice and the secondary electron created by its first ionization will also ionize once so that $P_4(z) = 1$ at this point. In the region $0.13$ µm $< z < 0.17$ µm multiplications of 5, 6 and 7 become possible with each probability $P_m(z)$ peaking at slightly higher $z$ as $m$ increases. $m = 8$ for the majority of electrons injected into the multiplication region at $z = W$. In this case the injected electron ionizes three times, the first secondary electron created ionizes twice (with a further ionization due to one of its secondaries) and the next secondary electron created ionizes once giving a multiplication of 8.

Figure 4 shows $P_m(W)$ plotted against $m$ for $M(W) \approx 8$ and for ionization coefficient ratios $k = 0$ (+), $k = 0.1$ (×) and $k = 1$ (o). The parameters used for $\alpha(z)$ are the same as those used for Figure 3 and $\beta(z) = \alpha(z)$ for $k = 1$, $\beta(z) = 0$ for $k = 0$ and $\beta(z)$ is given by $a = 2.2892$ µm$^{-1}$, $b = 300$ µm$^{-1}$ and $l = 0.0415$ µm for $k = 0.1$. The results are normalised to a mean multiplication of 8 by selecting the multiplication width as shown in Table 1. When $k = 0$, $P_m(W)$ is strongly peaked at $m = 8$ with $P_7(W) \approx P_9(W) < 0.01$ and $P_6(W) \approx P_{10}(W) < 0.002$. For $k = 0.1$ the behaviour is similar to the low field behaviour shown in Figure 2 except that $P_m(W) \approx 0$ for $m < 4$. For $k = 1$, $P_m(W)$ is significantly greater if $m$ is even for $m < 16$. In this case $W = 0.0773$ µm so that a carrier can ionize once only as it crosses the multiplication region ($W < 2l$). Therefore a multiplication greater than 2 can only be realised by a chain of alternate electron initiated and hole initiated ionizations. An electron injected at $z = W$ will ionize at $z \approx 0.03$ μm and the secondary hole may leave the multiplication region before ionizing resulting in a high probability for $m = 2$. If the hole does ionize it will create an electron close to $z = W$. The probability of this secondary electron surviving without ionization before it reaches $z = 0$ is small resulting in a low probability for $m = 3$ and a much higher probability for $m = 4$. This effect persists for higher $m$ but becomes less significant. Hole initiated ionization occurs further away from $z = W$ as $m$ increases because ionization chains that result in electrons created close to $z = W$ are more likely to be removed from the population.

4. Excess noise factor

The excess noise factor, $F(z)$, for an e-h pair created at $z$ is given by the following equation

$$F(z) = \frac{\langle m^2 \rangle}{\langle m \rangle^2} \tag{24}$$

where $\langle m \rangle$ is the mean multiplication equal to $M(z)$ given by either equation (6) or (23) and $\langle m^2 \rangle$ is the mean square multiplication given by the equation (25).

$$\langle m^2 \rangle = \sum_{m=1}^{N} m^2 P_m(z) \qquad N \to \infty \tag{25}$$

Figure 5 shows the excess noise factor, $F(W)$, against multiplication, $M(W)$, for injected electrons and for five different values of $k$, the ionization coefficient ratio. The parameters used for $\alpha(z)$ are $a = 1$ μm$^{-1}$, $b = 10$ μm$^{-1}$ and $l = 0.15$ μm and the parameters used for $\beta(z)$ are $b = 10$ μm$^{-1}$ and $l = 0.15$ μm with $a$ selected to give the required $k$ as determined by equation (7). These parameters are the same as those used for the calculation of multiplication as shown in Figure 6 of the previous paper [12] and are appropriate for the 'low field' non local ionization coefficient shown in Figure 2 of that paper [12]. Different multiplications, $M(W)$, are found by changing the multiplication region width, $W$. The solid lines show $F(W)$ calculated using equation (24) and the dashed lines show the excess noise factor calculated using the standard McIntyre [1] equation (1) assuming local ionization coefficients. Clearly the inclusion of non local effects reduces the excess noise factor and this is consistent with previous works [3-7, 11] that have studied the effect of a dead space on avalanche noise. The conclusion is that the dead space acts to reduce both avalanche multiplication and noise so that the excess noise factor is decreased for a given multiplication.

Now consider the excess noise factor at the higher field value as shown in Figure 6. In this case $\alpha(z)$ is given by $a = b = 300$ µm$^{-1}$ and $l = 0.0415$ µm and $\beta(z)$ is given by $b = 300$ µm$^{-1}$ and $l = 0.0415$ µm with the value of $a$ (shown in Table 1) selected to give the required $k$ as determined by equation (7). These parameters give $\alpha(z)$ as shown in Figure 3 of the previous paper [12] and $\beta(z)$ is determined by the method used to calculate $M(W)$ as shown in Figure 9 of that paper [12]. The multiplication region width, $W$, is used to give different multiplication values, $M(W)$. The solid lines represent $F(W)$ calculated using equation (24) and the dashed lines represent the excess noise factor predicted by equation (1). Again the result at a high field is significantly different from the lower field result. The excess noise factor is no longer a monotonically increasing function of multiplication, $M(W)$. There are two key differences: firstly the excess noise factor reduces to nearly 1 for multiplications of 2, 4, 8 and 16 depending on the value of $k$ and secondly for multiplications greater than 14 the excess noise factor does not increase monotonically with $k$.

The first of these two differences is the easiest to explain. Figure 4 shows that for $k = 0$ and $M(W) \approx 8$ the gain is essential deterministic with $P_8(W) \approx 0.98$ and therefore the excess noise factor must be nearly one. This "nearly" deterministic behaviour also occurs at $M(W) = 2$ for $k \leq 0.3$, at $M(W) = 4$ for $k \leq 0.03$ and at $M(W) = 16$ if $k = 0$. By comparison with Figure 9 of the previous paper [12] it can be seen that the excess noise factor, $F(W)$, decreases to nearly 1, the value expected for a totally noiseless process, when the multiplication reaches a level section. In general (using the non local ionization coefficients appropriate for the higher field value) if the multiplication changes slowly with increasing multiplication width (or voltage) then the excess noise factor should reduce to near one. This non local effect is in addition to the dead space effect observed in Figure 5 and in previous work [3-7, 11].

The second difference is more difficult to explain and would seem, at first, paradoxical. Figure 7 shows the excess noise factor, $F(W)$, against ionization coefficient ratio, $k$. The two dashed lines show the relationship predicted by equation (1) for $M(W) = 16$ (upper line) and $M(W) = 8$ (lower line) using the local ionization coefficient approximation. The triangles and diamonds show the excess noise factor for $M(W) \approx 8$ and $M(W) \approx 16$ respectively calculated using non local ionzation coefficients appropriate for the lower field value (open symbols) and the higher field value (closed symbols). The parameters used were the same as those used to find Figure 5 for the low field case and Figure 6 for the high field case with the parameter *a* for holes used to select the required $k$ and the multiplication region width $W$ used to get the required multiplication, $M(W)$. Selected values are given in Table 1. At the lower field value the excess noise factor, $F(W)$, increases monotonically with $k$ as shown by the open symbols in Figure 7. However this is not the case at the high field value where the excess noise factor peaks at $k = 0.075$ and $k = 0.3$ for $M(W) \approx 16$ so that the excess noise factor is greater for $k = 0.3$ than for $k = 1$ as shown in Figure 6. Similarly for $M(W) \approx 8$ the excess noise factor peaks at $k = 0.22$ and $k = 0.98$. Furthermore the excess noise factor is greater for $M(W) \approx 8$ than it is for $M(W) \approx 16$ for ionization coefficient ratios in the range, $0.94 < k < 0.99$.

In order to explain these seemingly paradoxical results it is necessary to plot the excess noise factor, $F(W)$, against multiplication region width, $W$. Figure 8 shows $F(W)$ plotted against $W/l$ where $l$ is the length of the dead space region for $k = 1$ (top), 0.3, 0.1, 0.03 and 0 (bottom) shown by the solid lines and for $M(W) \approx 8$ (triangles) and $M(W) \approx 16$ (diamonds). The excess noise factor peaks at multiplication region widths slightly greater than an integral number of dead spaces. First consider the excess noise factor for $k = 0$. For multiplication region widths of less than one dead space the multiplication is 1 with certainty so that there is no avalanche noise and the excess noise factor is 1. When the multiplication region width, $W$, is slightly greater than the dead space width both $m = 1$ and $m = 2$ are possible and the excess noise factor increases to reflect this. In the interval $1.5l < W < 2l$ for $k = 0$, $m = 2$ with near certainty and the excess noise factor falls back to 1 as a result. This pattern repeats so that as the multiplication region width increases by an additional dead space the multiplication doubles (as shown in Figure 9 of [12]) and the excess noise factor increases as the multiplication makes that transition.

Now consider the peaks in $F(W)$ for a constant multiplication. In particular consider the peak for $M(W) \approx 16$ at $W = 3.2l$. In the interval $2.5l < W < 3.2l$ the injected electron will ionize twice and one of it's secondary electrons will ionize once with near certainty. A further 12 ionizations (on the average) are therefore required for a mean multiplication of 16 and these ionizations must be initiated firstly by holes and then by secondary electrons created as a result of hole initiated ionization. The multiplication due to the injected electron is nearly deterministic so that the excess noise factor will depend only on the hole initiated ionization and subsequent secondary electrons. This excess noise factor can be expected to increase with increasing multiplication region width, $W$, and decrease with decreasing $k$ and $\beta(z)$. Although $k$ must decrease as $W$ increases in order to keep the multiplication constant, the width effect dominants so that the excess noise factor increases with $W$ in this interval. Now consider the interval $3.2l < W < 3.6l$. In this interval the multiplication due to the injected electron and it's secondary electrons increases from 4 to 8 so that the number of additional ionizations initiated by holes and subsequent secondary electrons decreases from 12 to 8 (on the average). At the same time $k$ and $\beta(z)$ are decreasing in order to keep the multiplication constant and the combined effect of decreasing $k$ and decreasing hole initiated ionization dominates the effect of increasing $W$ so that the excess noise factor decreases in this interval.

It should be noted that for a constant multiplication region width, $W$, the multiplication and the excess noise factor increases monotonically with $k$. Best performance in a real device with a fixed width would be achieved by minimising $k$.

## 5. Conclusion

A method for calculating the probability, $P_m(z)$, of an electron hole pair created at $z$ resulting in a multiplication of $m$ has been derived. $P_m(z)$ has been evaluated for the two different field values and at the lower field value the results are similar to that expected for local ionization coefficients. Results for $P_m(z)$ at the higher field value show a more deterministic behaviour; especially for ionization initiated by only one carrier type. The excess noise factor has been calculated using non local ionization coefficients and the results show two effects, one attributed to the dead space and the other attributed to the 'resonance' effect. The dead space effect can be seen at both field values and it acts to reduce the excess noise factor to a value less than that predicted by McIntyre [1]. The resonance effect is observed at the higher field only and it acts to further reduce the excess noise factor to nearly one when the multiplication curves are level for $M$ = 2, 4, 8 and 16.

Acknowledgements

I thank Professor R. C. Woods for a critical reading of this manuscript.

# References.

Table Captions.

Table 1  Selected values of parameters used in the calculations and results for the higher field value. The second column gives the value of the *a* parameter for holes. For electrons $a = 300$ μm$^{-1}$. For both carriers $b = 300$ μm$^{-1}$ and $l = 0.0415$ μm.

Figure Captions.

Fig. 1   $P_m(z)$ for $1 \leq m \leq 28$, $k = 0$ and an electric field of $3 \times 10^7$ Vm$^{-1}$ with dashed lines for even $m$ and solid lines for odd $m$ with $m$ increasing from left to right.

Fig. 2   $P_m(W)$ against $m$ for $M(W) \approx 8$ and an electric field of $3 \times 10^7$ Vm$^{-1}$ with $k = 0$ (+), 0.1 (×) and 1 (o).

Fig. 3   $P_m(z)$ for $1 \leq m \leq 8$, $k = 0$ and an electric field of $10^8$ Vm$^{-1}$ with dashed lines for even $m$ and solid lines for odd $m$.

Fig. 4   $P_m(W)$ against $m$ for $M(W) \approx 8$ and an electric field of $10^8$ Vm$^{-1}$ with $k = 0$ (+), 0.1 (×) and 1 (o).

Fig. 5   Excess noise factor, $F(W)$, for injected electrons against multiplication, $M(W)$, for an electric field of $3 \times 10^7$ Vm$^{-1}$ calculated using non local ionization coefficients (solid lines) and McIntyres equation (1) (dashed lines) for $k = 1$, 0.3, 0.1, 0.03 and 0 with $k$ increasing from bottom to top for both sets of curves.

Fig. 6   Excess noise factor, $F(W)$, for injected electrons against multiplication, $M(W)$, for an electric field of $10^8$ Vm$^{-1}$ calculated using non local ionization coefficients (solid lines) and McIntyres equation (1) (dashed lines) for $k = 1$, 0.3, 0.1, 0.03 and 0 with $k$ increasing from bottom to top for both sets of curves.

Fig. 7   Excess noise factor, $F(W)$, for injected electrons against ionization coefficient ratio, $k$, for $M(W) \approx 8$ (triangles) and $M(W) \approx 16$ (diamonds) and for low field values (open symbols) and high field values (closed symbols). Dashed lines show values predicted by McIntyre's equation (1).

Fig. 8   Excess noise factor, $F(W)$, for injected electrons against multiplication region width, $W$, for an electric field of $10^8$ Vm$^{-1}$ calculated using non local ionization coefficients for $k$ = 1, 0.3, 0.1, 0.03 and 0 (solid lines) and for $M(W) \approx 8$ (triangles) and $M(W) \approx 16$ (diamonds).

| $k$ | $a$ ($\mu m^{-1}$) | $W$ ($\mu m$) | $M(W)$ Eq.(6) | Number of iterations of Eq.(6) | $M(W)$ Eq.(23) | $N$ Eq.(22) | $F(W)$ |
|---|---|---|---|---|---|---|---|
| 1 | 300 | 0.0894 | 16.184 | 283 | 16.178 | 328 | 3.073 |
| 0.6 | 28.213 | 0.1022 | 15.928 | 87 | 15.922 | 329 | 2.236 |
| 0.3 | 8.6414 | 0.1303 | 16.058 | 116 | 16.049 | 655 | 3.690 |
| 0.1 | 2.2892 | 0.1570 | 16.100 | 48 | 16.095 | 214 | 1.637 |
| 0.075 | 1.6740 | 0.1700 | 15.991 | 50 | 15.986 | 242 | 1.725 |
| 0.03 | 0.6407 | 0.189 | 15.671 | 30 | 15.667 | 123 | 1.295 |
| 0 | - | 0.22 | 16.037 | 6 | 16.071 | 25 | 1.004 |
| 1 | 300 | 0.0773 | 8.002 | 152 | 8.001 | 159 | 2.853 |
| 0.98 | 231.7 | 0.0867 | 8.025 | 193 | 8.023 | 201 | 3.285 |
| 0.4 | 13.2304 | 0.1044 | 8.020 | 38 | 8.018 | 94 | 1.542 |
| 0.22 | 5.7436 | 0.1296 | 8.001 | 48 | 7.999 | 148 | 1.883 |
| 0.1 | 2.2892 | 0.1408 | 8.021 | 26 | 8.019 | 69 | 1.375 |
| 0 | - | 0.173 | 7.996 | 5 | 7.995 | 12 | 1.000 |

Table 1.

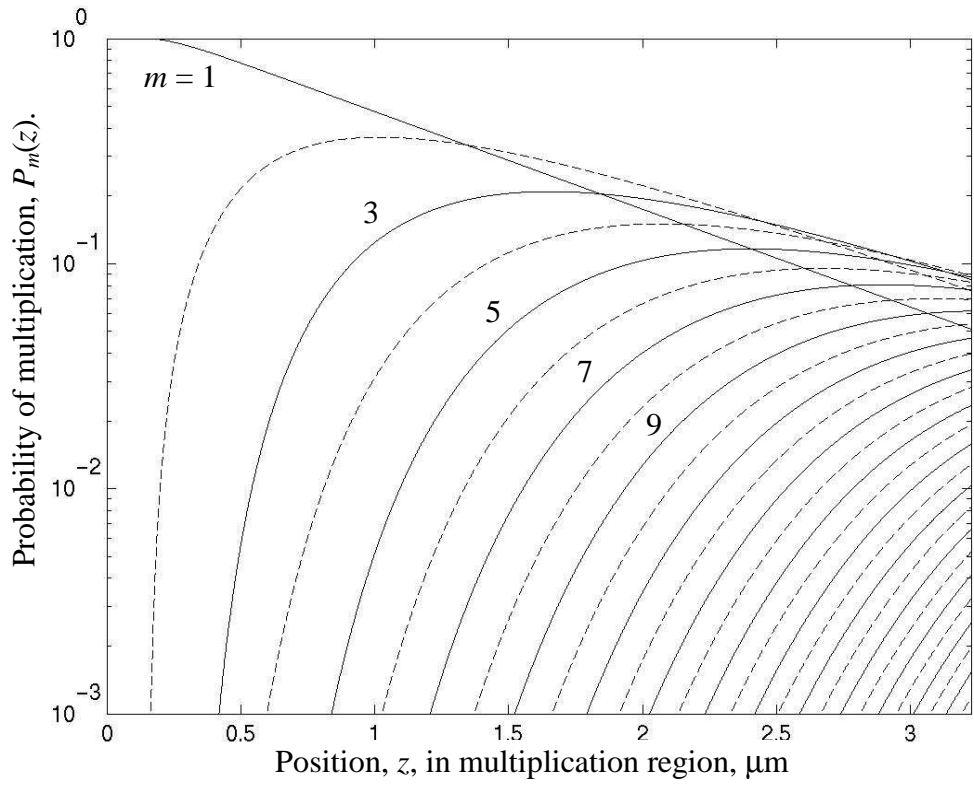

Figure 1.

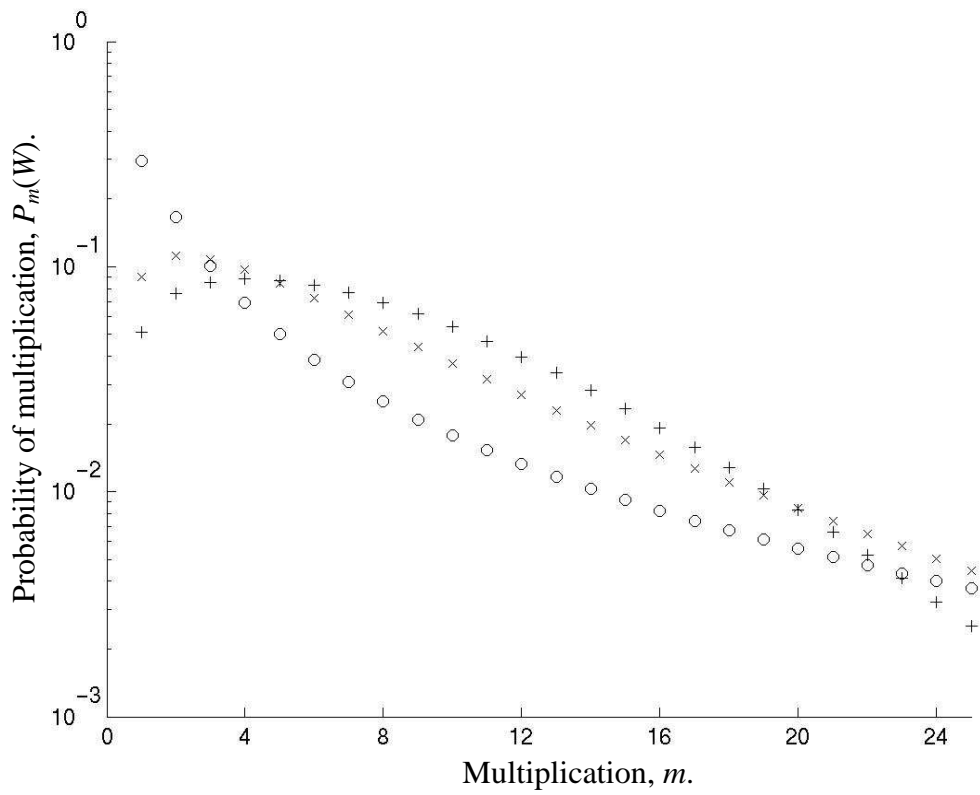

Figure 2.

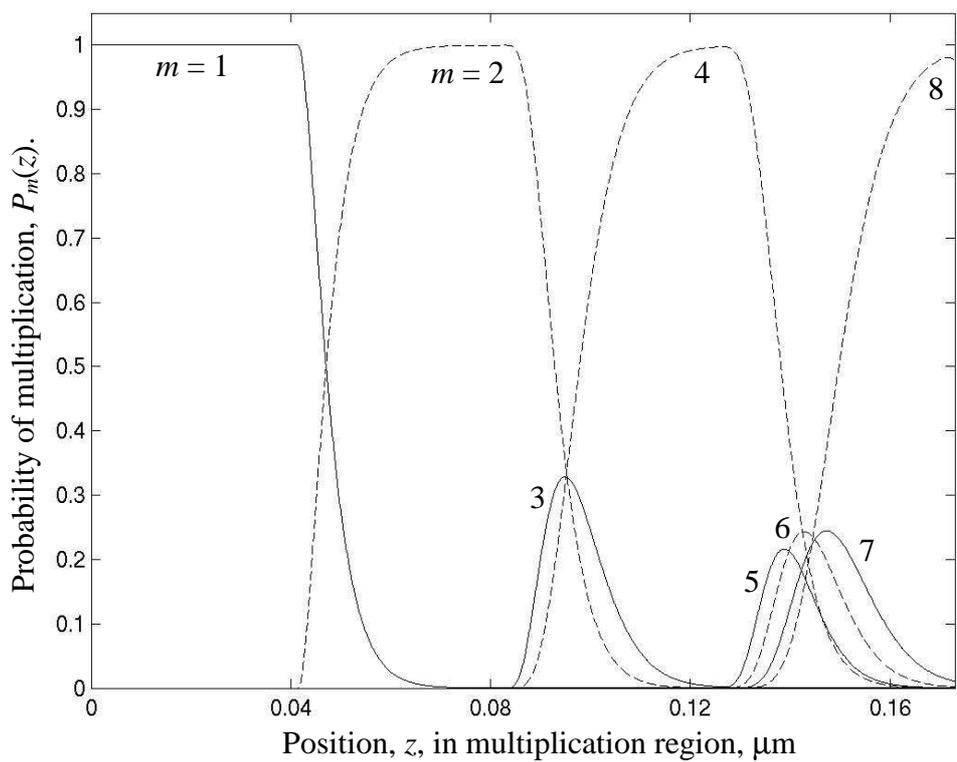

Figure 3.

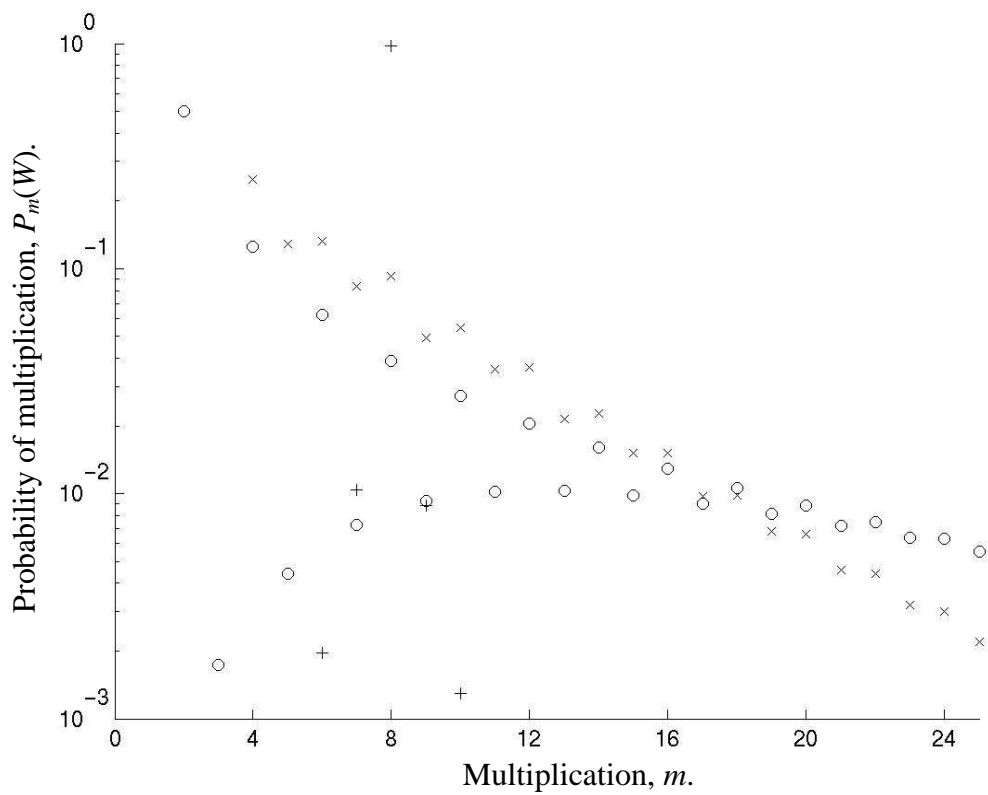

Figure 4.

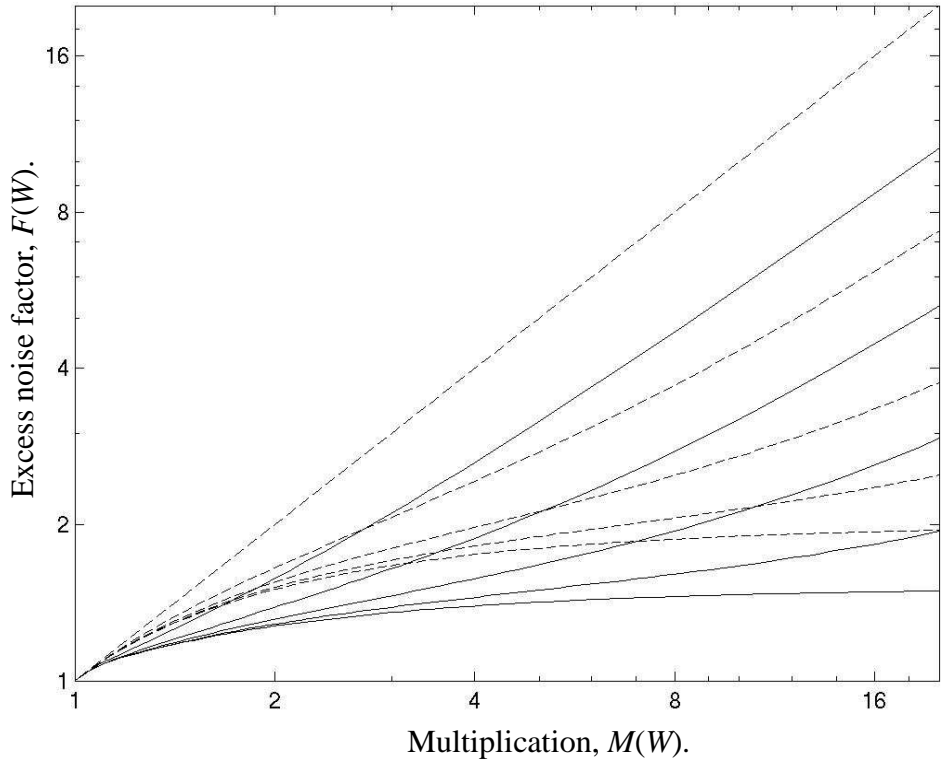

Figure 5.

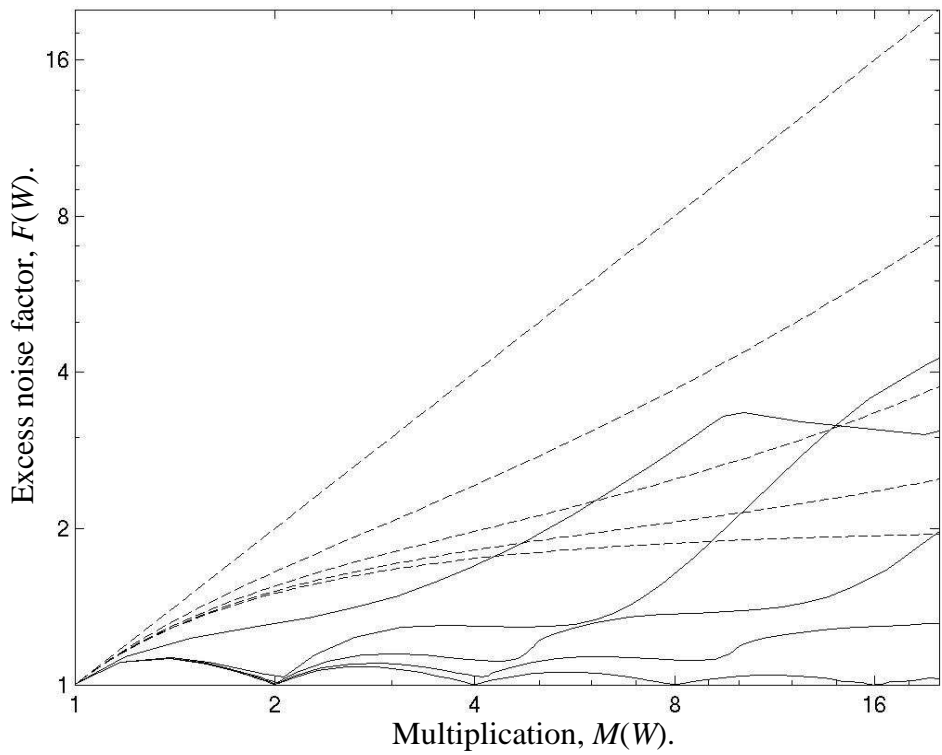

Figure 6.

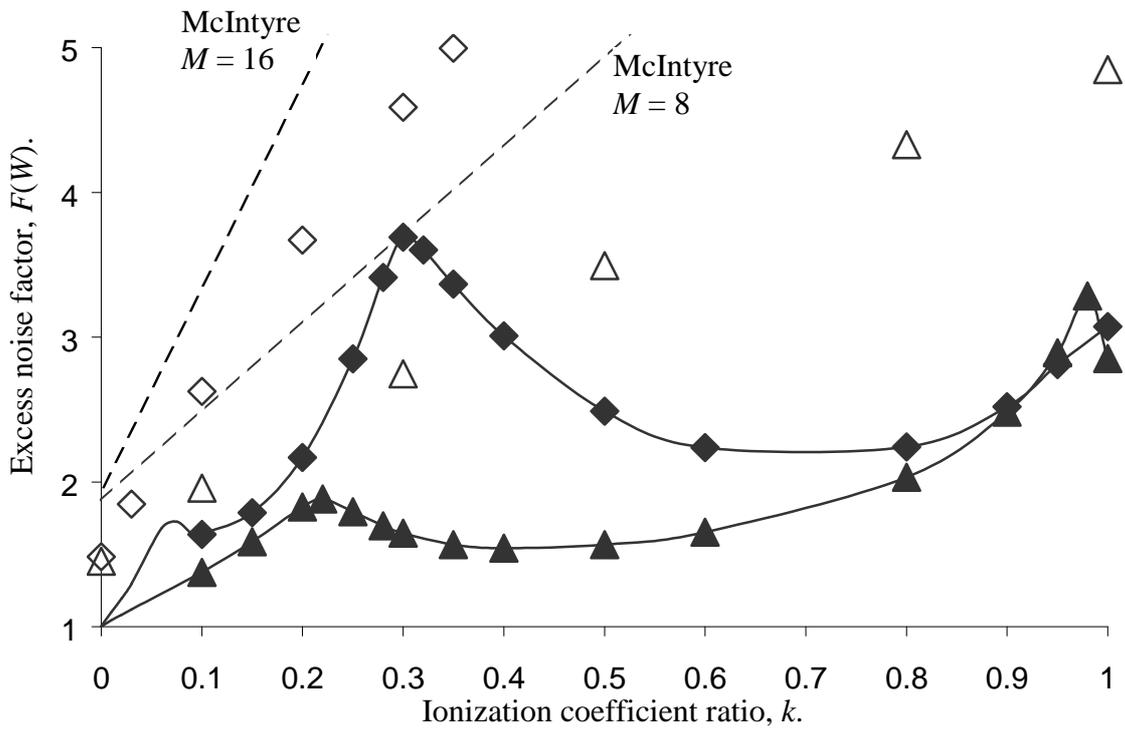

Figure 7.

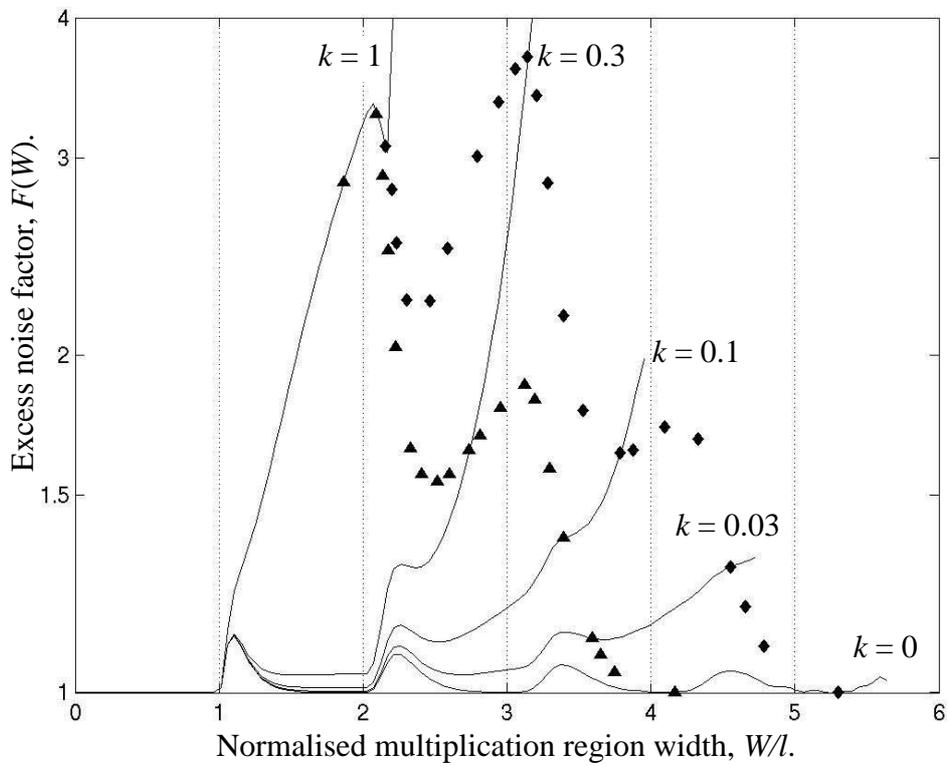

Figure 8.